\documentclass[twocolumn,aps,pre,showpacs]{revtex4}

\usepackage{graphicx}
\begin{document}

\title{Four-point susceptibility of a glass-forming binary mixture: 
Brownian dynamics}
\author{Grzegorz Szamel and Elijah Flenner}
\affiliation{Department of Chemistry,
Colorado State University, Fort Collins, CO 80523}

\date{\today}


\begin{abstract}
We study the four-point dynamic susceptibility $\chi_4(t)$ 
obtained from Brownian dynamics computer simulations 
of the Kob-Andersen Lennard-Jones mixture.
We compare the results of the simulations with 
qualitative predictions of the mode-coupling theory. In addition, we
test an estimate of the four-point susceptibility recently proposed
by Berthier \textit{et al.} [Science, \textbf{310}, 1797 (2005)]. 
\end{abstract}
\maketitle

\section{\label{intro}Introduction}

The origin of the extreme slowing down of liquids' dynamics upon approaching 
the glass transition and the very nature of this transition are still
hotly debated despite many experimental, simulational, and theoretical 
studies performed in the last two decades \cite{DebStil}.
One of the fundamental difficulties
is that the slowing down appeared to be a local, small
scale phenomenon which is not accompanied by a growing correlation length.
No long-range correlations have ever been found in any static quantity upon 
approaching the glass transition. The attention has 
recently shifted to dynamic correlations \cite{BB}, and there is evidence that 
there is a dynamic correlation length that slowly grows upon
approaching the glass transition. Unfortunately this dynamic correlation 
length cannot be easily obtained from experimental data.

The dynamic correlation length is often obtained
from the four-point dynamic susceptibility $\chi_4(t)$. This  
susceptibility is related to a space integral of a four-point density 
correlation function that quantifies correlations between relaxation 
processes at different points in space. It is assumed that 
the increase of the dynamic susceptibility signals the growth of the 
range of the correlations between relaxation processes. 
Furthermore, the four-point dynamic susceptibility can be 
used to discriminate between different theoretical approaches to 
glassy dynamics. For example, in a recent paper \cite{Toninelli4p} 
Toninelli \textit{et al.} discussed predictions for $\chi_4(t)$ 
obtained from various theoretical approaches to glassy dynamics and compared 
these predictions with two different simulations of 
atomistic models of glass forming liquids. 

A problem with the four-point susceptibility is that it is  
difficult to extract from experiments. This problem has been addressed
in another recent paper \cite{BerthierScience}. 
Berthier \textit{et al.}~argued that the four-point susceptibility can be
estimated using derivatives of an intermediate scattering function
with respect to thermodynamic parameters such as temperature or density. 
Since the intermediate scattering function can be easily obtained from 
experiments, Ref.~\cite{BerthierScience} introduced a way to experimentally 
investigate the existence of a growing correlation length. 

The goal of this contribution is twofold. First, we analyze 
the four-point susceptibility of the Kob-Andersen
\cite{KobAndersen} Lennard-Jones binary mixture undergoing
Brownian dynamics. Second, we test the approximate
estimate for the four point susceptibility proposed by Berthier \textit{et al.}
\cite{BerthierScience}. 

The paper is organized as follows: in Sec. \ref{model} we briefly review
the details of the simulation; in Sec. \ref{chi4} we define the 
four-point susceptibility and analyze the simulation results for this
quantity; in Sec. \ref{estimate} we compare the four-point susceptibility 
obtained directly from computer simulations 
to the approximate expression derived by Berthier \textit{et al.};
and in Sec. \ref{conclusion} we discuss our findings.

\section{\label{model} Simulation details} 

We simulated a binary mixture that was introduced by
Kob and Andersen which consists of $N_A=800$ particles of type A and
$N_B=200$ particles of type B. The interaction potential is
$V_{\alpha \beta}(r) = 4\epsilon_{\alpha \beta}[
({\sigma_{\alpha \beta}}/{r})^{12} - ({\sigma_{\alpha \beta}}/{r})^6]$,
where $\alpha, \beta \in \{A,B\}$, $\epsilon_{AA} = 1.0$,
$\sigma_{AA} = 1.0$, $\epsilon_{AB} = 1.5$, $\sigma_{AB} = 0.8$,
$\epsilon_{BB} = 0.5$, and $\sigma_{BB} = 0.88$.  The interaction potential 
was cut off 
at $2.5\ \sigma_{\alpha \beta}$. We used a cubic simulation cell with 
the box length of $9.4\ \sigma_{AA}$ with periodic boundary conditions.

We performed Brownian dynamics simulations.  The equation of
motion for the position of the $i_{th}$ particle of type $\alpha$, 
$\vec{r}\,_{i}^{\alpha}$, is
\begin{equation}\label{Lang}
\dot{\vec{r}}\,_{i}^{\alpha} = \frac{1}{\xi_0} \vec{F}_i^{\alpha}
+ \vec{\eta}_i(t) ,
\end{equation}
where $\xi_0$ is the friction coefficient of an isolated particle, 
$\xi_0 = 1.0$, and $\vec{F}_i^{\alpha}$
is the force acting on the $i_{th}$ particle of type $\alpha$,
\begin{equation}\label{force}
\vec{F}_i^{\alpha}= - \nabla_i^{\alpha} \sum_{j} \sum_{\beta=1}^2
V_{\alpha \beta}\left(\left|\vec{r}\,_i^{\alpha} - 
\vec{r}\,_j^{\beta} \right| \right)
\end{equation}
with $\nabla_i^{\alpha}$ being the gradient operator with 
respect to $\vec{r}\,_i^{\alpha}$ (note that the term with 
$\beta=\alpha$ and $i=j$ has to be excluded from the double sum in 
Eq. (\ref{force})). 
In Eq. (\ref{Lang})
the random noise $\vec{\eta}_i$ satisfies the fluctuation-dissipation
theorem,
\begin{equation}\label{fd}
\left\langle \vec{\eta}_i(t) \vec{\eta}_j(t') \right\rangle =
2 D_0 \delta(t-t') \delta_{ij} \mathbf{1}.
\end{equation}
Here $D_0$ is the diffusion coefficient of an isolated 
particle, $D_0 = k_B T/\xi_0$, where
$k_B$ is Boltzmann's constant. Furthermore, in Eq. (\ref{fd}) $\mathbf{1}$ is
the unit tensor.  The equations of motion (\ref{Lang}-\ref{fd}) 
allow for diffusive motion of the center of mass, thus 
all the results will be presented relative to the
center of mass (\textit{i.e.}\ momentary positions of all the particles are 
always relative to the momentary position of the center of mass \cite{remark}).
The results are presented in terms of the reduced units
with $\sigma_{AA}$, $\epsilon_{AA}$, $\epsilon_{AA}/k_B$,
and $\sigma_{AA}^2 \xi_0/\epsilon_{AA}$ being the units of length, energy,
temperature, and time, respectively. In these units the short-time 
self-diffusion coefficient is proportional to the temperature, thus
the time is rescaled to a reduced time equal to $t D_0/\sigma_{AA}^2$
to facilitate comparisons with theoretical approaches that often
assume temperature-independent short-time dynamics.

The equations of motion were solved using a Heun algorithm 
with a small time step of $5 \times 10^{-5}$.  We simulated a broad range of
temperatures $0.44 \le T \le 5.0$. Here we present results for the 
following temperatures: $T = 0.45$, 0.47, 0.50, 0.55, 0.60, 0.65, 0.80, and 1.0.
We ran equilibration runs and 4-6 production runs.
The equilibration runs were typically twice to four times 
shorter than the production runs, and the latter
were up to $1.2 \times 10^9$ time steps long for the lowest temperature
discussed here, $T=0.45$. 
The results presented are averages over the production runs.

\section{\label{chi4}Four-point dynamic susceptibility}

In the context of the dynamics of 
supercooled liquids, the four-point susceptibility has been 
introduced by Glotzer and collaborators \cite{Sharon}.
The four-point susceptibility that we discuss in this paper is 
slightly different from that defined in Refs.~\cite{Sharon}. 
We consider here the susceptibility which was extensively  
analyzed by Toninelli \textit{et al.}~\cite{Toninelli4p}, which 
is defined as the variance
of the fluctuations of the self-intermediate scattering function.

We start with the definition of the self-intermediate scattering
function, $F_s(k;t)$,
\begin{equation}\label{Fs}
F_s(k;t) = \left<  \frac{1}{N} \sum_i \cos \vec{k}\cdot 
[\vec{r}_i(t) - \vec{r}_i(0)] \right>.
\end{equation}
Next, we define the fluctuation of the instantaneous value of the 
scattering function, $\delta F_s(\vec{k};t)$,
\begin{equation}\label{deltaFs}
\delta F_s(\vec{k};t) = \frac{1}{N} \sum_i \cos \vec{k}\cdot 
[\vec{r}_i(t) - \vec{r}_i(0)] - F_s(k;t).
\end{equation}
The four-point susceptibility, $\chi_4(t)$, is then defined as
\begin{equation}\label{chi4b}
\chi_4(t) = 
N \left< \delta F_s(\vec{k};t) \delta F_s(\vec{k};t) \right>.
\end{equation}
The four-point susceptibility depends on the
wave vector $\vec{k}$ (and it is sometimes denoted by 
$\chi_{\vec{k}}(t)$). This wave vector is customarily fixed at the position of
the maximum of the static structure factor \cite{Chandler}.

The system considered in this paper is a two-component mixture, thus
instead of one self-intermediate scattering function $F_s(k;t)$ 
we could introduce
two different scattering functions involving particles $A$ and $B$. All the
results presented in this paper concern the $A$ particles only, thus,
\textit{e.g.} sums in Eqs. (\ref{Fs}-\ref{deltaFs}) run only over the
$A$ particles. Since we 
are not presenting any results for the $B$ particles we do not introduce 
additional sub-scripts or super-scripts indicating particle labels. The magnitude
of the wave vector
$\vec{k}$ is fixed at the  position of the maximum of the partial structure 
factor of the $A$ particles, $|\vec{k}|= 7.25$.

The mode coupling theory of the glass transition 
was formulated by G\"otze and collaborators
for Newtonian systems \cite{Goetze}, and was later extended 
by Szamel and L\"owen to Brownian systems
\cite{SL}. The theory makes predictions 
for the self- and collective intermediate scattering functions. 
Biroli and Bouchaud \cite{BBEPL} recently 
argued that the mode coupling theory could
be understood as a dynamic mean-field theory, and that the usual
mode coupling equations are saddle point equations obtained from an
action functional. This new interpretation made it possible to calculate 
fluctuations of the order parameter (\textit{i.e.}\ fluctuations of a 
two-point dynamic correlation function) 
from the inverse of the second derivative of the action functional. The
details of the calculation have not been published, but the main predictions
have already been discussed and compared with Newtonian dynamics simulations
\cite{Toninelli4p}. According to Biroli and Bouchaud, on the $\beta$ 
relaxation time scale the four-point susceptibility grows with time 
as a power law, $\chi_4(t) \propto t^{\mu}$, with exponents equal to the 
standard mode coupling exponents $a$ and $b$ in the 
early and late $\beta$ regime, respectively. 
Furthermore,  $\chi_4(t)$ reaches its maximum value, $\chi_4(t^*)$, 
on the time scale of the $\alpha$ relaxation time, 
$t^* \sim \tau_{\alpha}$, and the maximum value diverges upon approaching
the mode coupling temperature, $\chi_4(t^*)\propto (T-T_c)^{-\gamma_1}$
with $\gamma_1=1$.

The authors of Refs.~\cite{Toninelli4p,BerthierScience} emphasized 
that these predictions are valid for Newtonian systems in the microcanonical
(NVE) ensemble and hinted that they may be different in the canonical
(NVT) ensemble. 
We present results obtained from Brownian dynamics
simulations. In such simulations constant temperature is maintained 
automatically by the equations of motion. Thus, in principle it is not
clear whether the predictions of Biroli and Bouchaud \cite{BBEPL}
are relevant for our simulation results. However, there are subtle theoretical
arguments that suggest that the reverse is true \cite{BBprivinfo}. 
In addition, at the level of the mode coupling equations, there is no
difference between Newtonian systems in the NVE or NVT ensemble, and
Brownian systems (beyond the microscopic time scale, \textit{i.e.}
on the time scales of the $\beta$ relaxation and longer) \cite{SL}. 
These reasons encouraged us
to compare our results obtained from Brownian dynamics
simulations to the predictions of Biroli and Bouchaud \cite{BBEPL}.

\begin{figure}
	\includegraphics[scale=0.25]{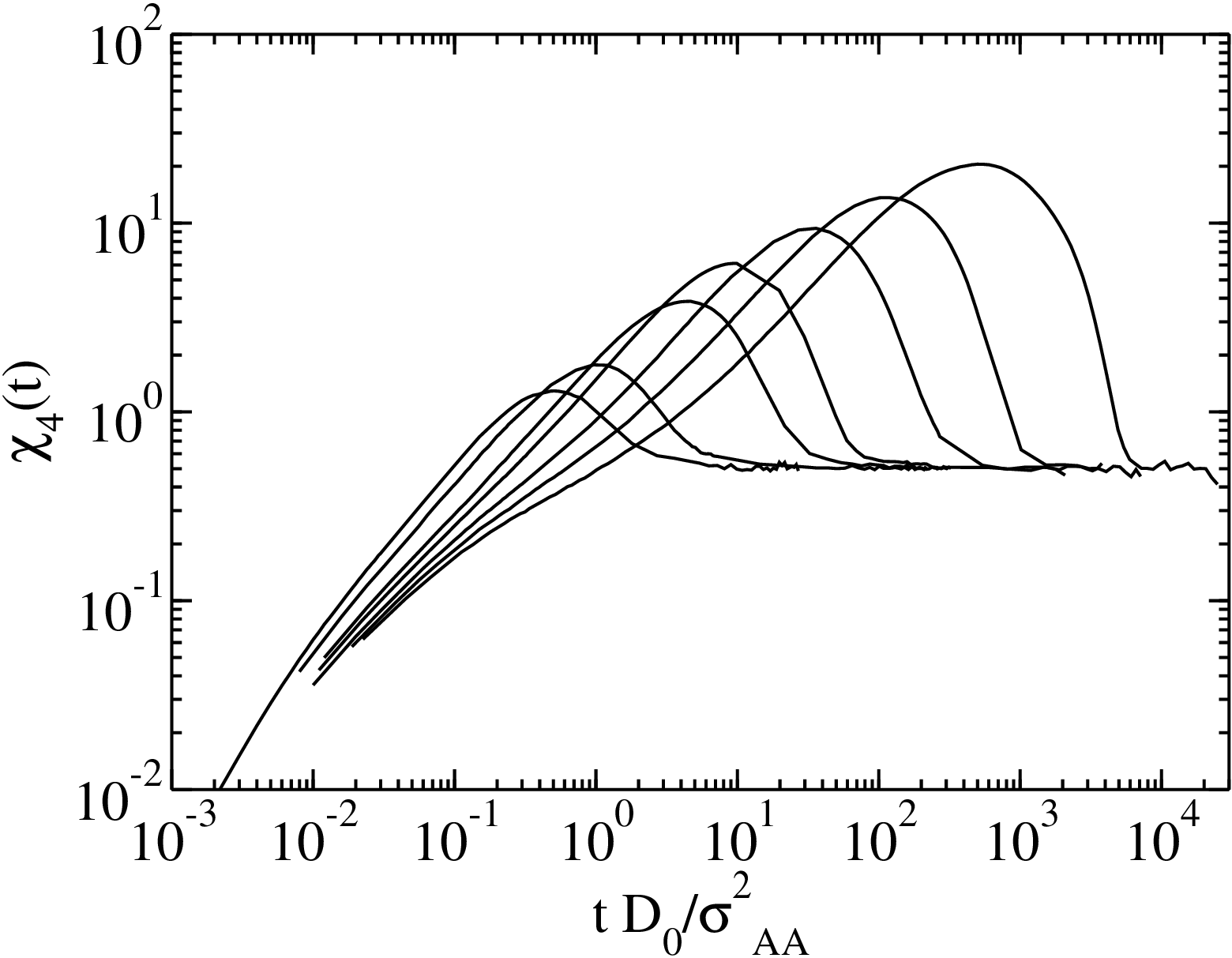}
\caption{\label{chi4fig} Time dependence of  the four-point susceptibility 
$\chi_4(t)$ plotted \textit{vs.} $t D_0/\sigma^2$ for $T=$1.0, 0.80,
0.60, 0.55, 0.50, 0.47, and 0.45 (left to right).}
\end{figure}

In Fig. \ref{chi4fig} we show the general shape of the four-point 
susceptibility, $\chi_4(t)$, for several temperatures. The time 
dependence of $\chi_4(t)$ is similar to that obtained for Newtonian
systems, see Figs. 4 and 5 of Ref. \cite{Toninelli4p}. In particular,
it can be argued that there is power law like growth in time of the four-point 
susceptibility as it approaches it's maximum value. 
This power law-like dependence 
will be further analyzed at the end of this section.

\begin{figure}
	\includegraphics[scale=0.25]{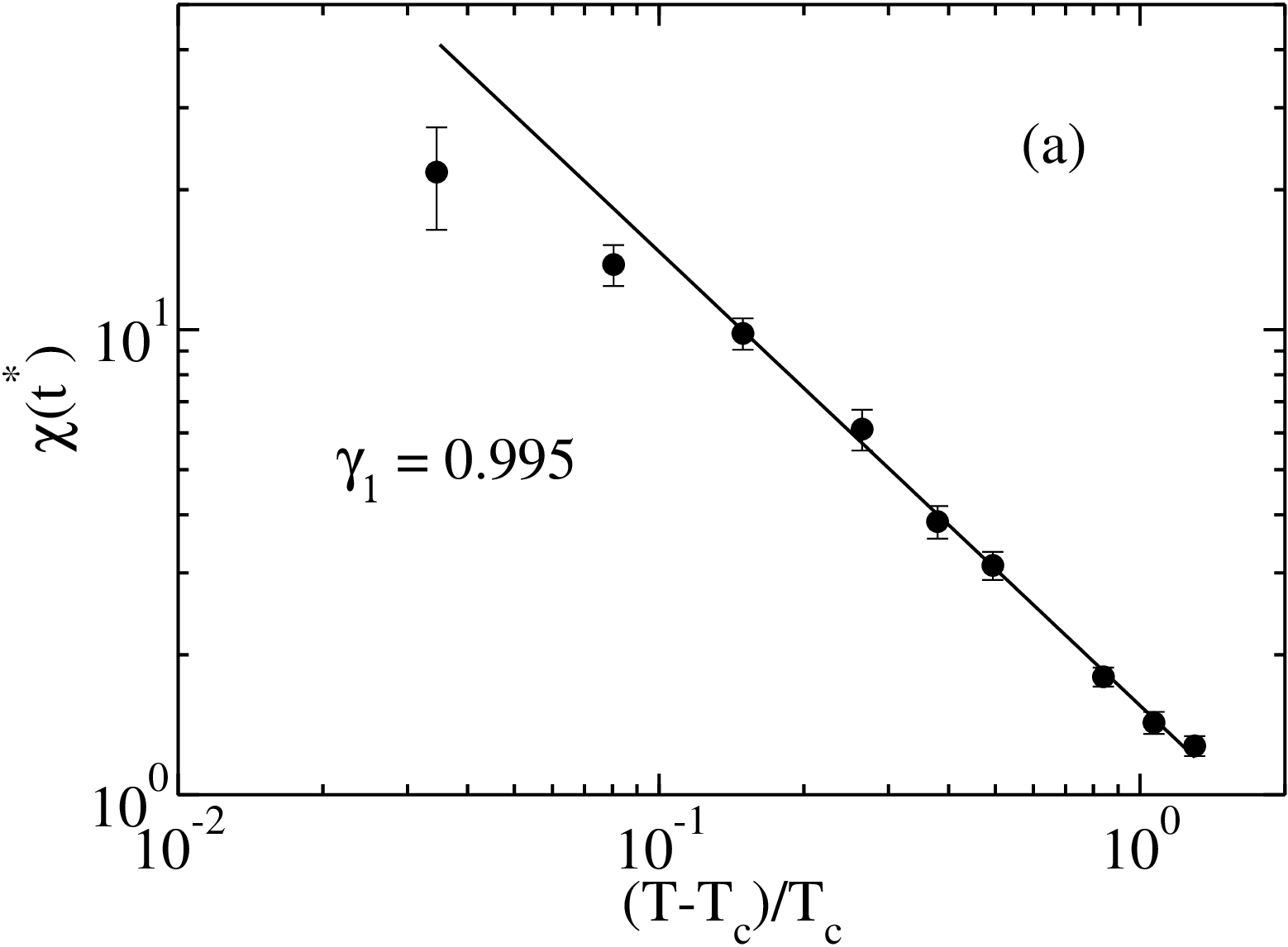}\\[0.25cm]
	\includegraphics[scale=0.25]{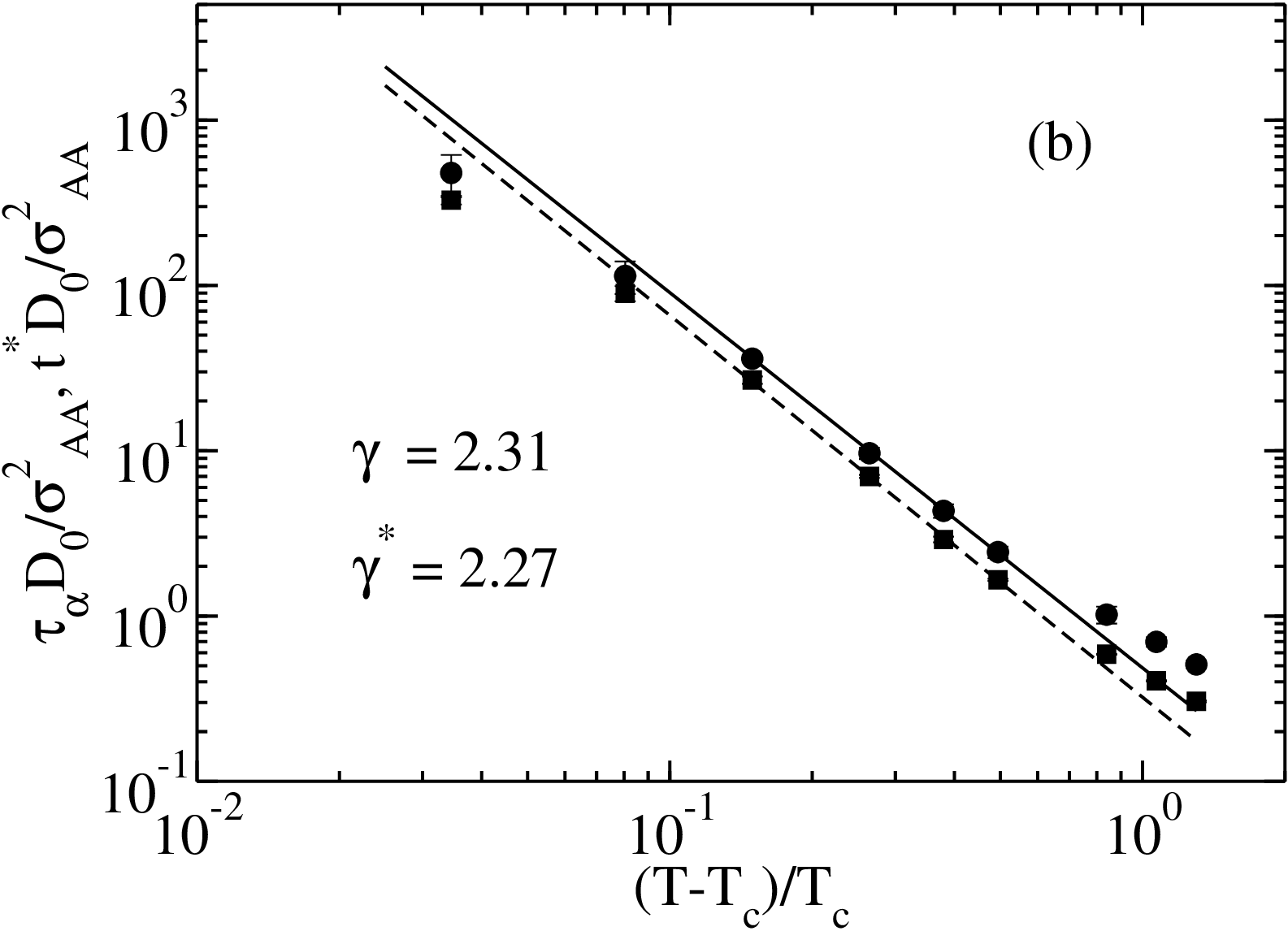}\\[0.25cm]
	\includegraphics[scale=0.25]{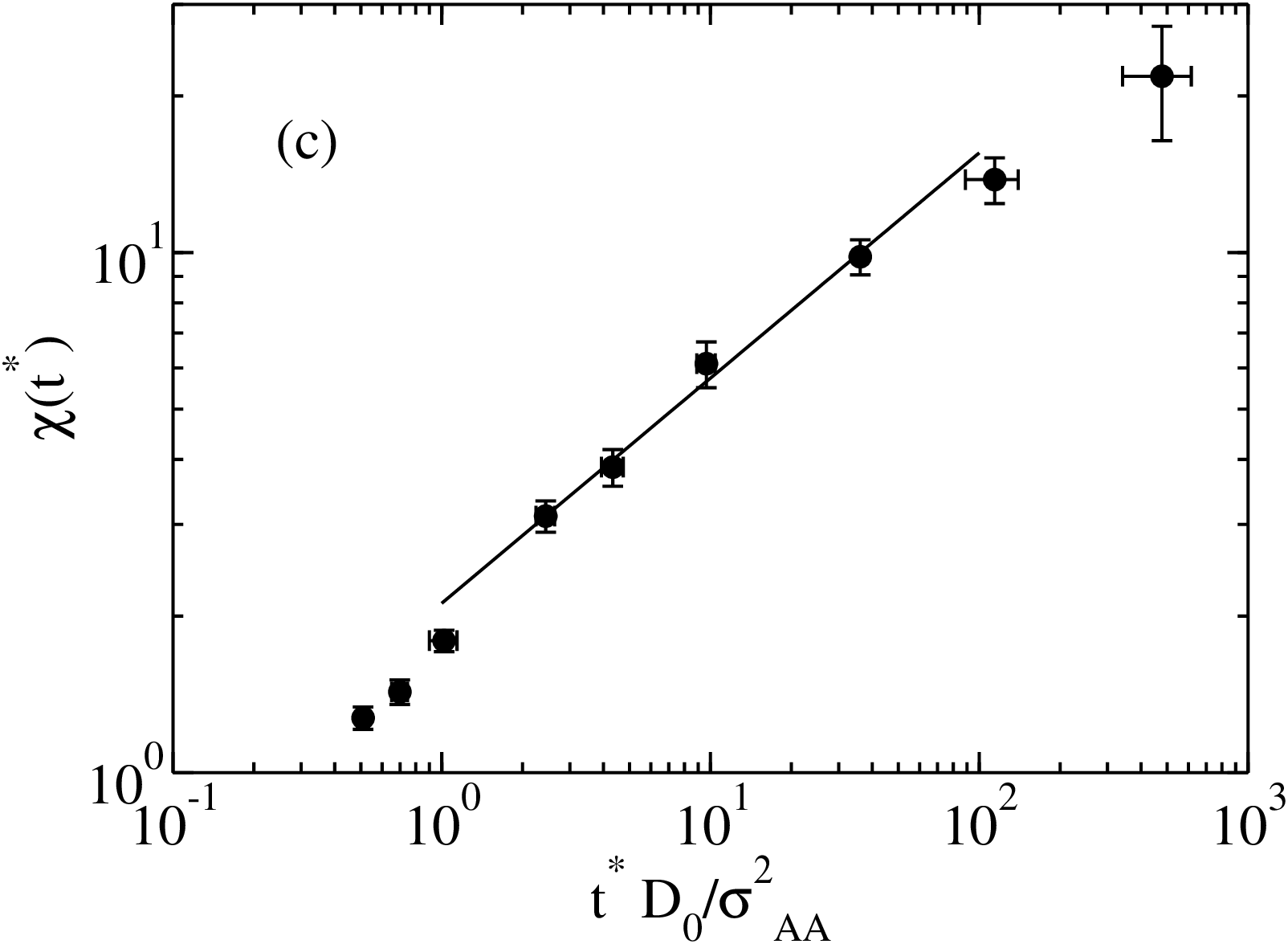}
\caption{\label{peaketc}
(a) Temperature dependence of the maximum value of the four-point 
susceptibility: $\chi_4(t^*)$ plotted \textit{vs.} $\epsilon=(T-T_c)/T_c$.
Solid line is the power law fit $\chi_4(t^*) \propto \epsilon^{-\gamma_1}$
with $\gamma_1$=0.995. 
(b) Temperature dependence of the time at which $\chi_4(t)$ reaches 
its maximum value, $t^*$ (circles), and the $\alpha$ relaxation time,
$\tau_{\alpha}$ (squares). Both times are plotted \textit{vs.} 
$\epsilon=(T-T_c)/T_c$. The solid line and the dashed line are power law fits
$t^*\propto \epsilon^{-\gamma^*}$ with $\gamma^*$=2.27 and 
$\tau_{\alpha}\propto \epsilon^{-\gamma}$ with $\gamma$=2.31 , respectively.
(c) The maximum value of the four-point 
susceptibility, $\chi_4(t^*)$, plotted \textit{vs.} $t^*$. The solid line is
the power law fit $\chi_4(t^*)\propto (t^*)^{1/\gamma_2}$ with $\gamma_2$ = 
2.31.}
\end{figure}

In Fig.~\ref{peaketc} we present the results of some quantitative analysis of $\chi_4(t)$. 
First, in Fig.~\ref{peaketc}a we show the temperature dependence of the
its maximum value, $\chi_4(t^*)$. We plot the maximum value \textit{vs.}
$\epsilon=(T-T_c)/T_c$ where $T_c=0.435$ is the standard value
of the mode coupling temperature for the Kob-Andersen Lennard-Jones binary 
mixture \cite{KobAndersen}. In an intermediate range
of temperatures, $\chi_4(t^*)$ grows as a power law 
with decreasing $(T-T_c)/T_c$, $\chi_4(t^*)\propto \epsilon^{-\gamma_1}$. 
The exponent obtained from the fit, 
$\gamma_1=0.995 \pm 0.05$, 
is very close to the theoretical prediction of Biroli 
and Bouchaud, $\gamma_1^{th}=1$. 
It should be noted that the range of reduced temperatures 
for which the power law dependence of $\chi_4(t^*)$ is observed coincides
with the range of reduced temperatures for which mode coupling theory was
found to correctly describe the time evolution of the self-intermediate
scattering function and the mean square displacement \cite{FS}.
In Fig.~\ref{peaketc}b we compare the temperature dependence of the 
time at which $\chi_4(t)$ reaches its maximum value, $t^*$, with that
of the $\alpha$ relaxation time, $\tau_{\alpha}$. The latter time
is defined as the time at which the self-intermediate 
scattering function decays to $e^{-1}$ of its initial value,
$F^s(q,\tau_{\alpha}) = e^{-1}$. We find that 
$t^*$ and $\tau_{\alpha}$ are very close and have the same temperature
dependence. In particular, in the same intermediate range
of temperatures these times grow according to power laws with decreasing 
$(T-T_c)/T_c$, $t^*\propto \epsilon^{-\gamma^*}$, 
$\tau_{\alpha}\propto \epsilon^{-\gamma}$.  
The exponents obtained from the fits, 
$\gamma^*=2.27 \pm 0.04$ and $\gamma=2.31 \pm 0.02$ are very close. The numerical solution
of the mode coupling equations predicts a slightly higher value of the 
scaling exponent for $\tau_{\alpha}$, $\gamma^{th} = 2.46$ \cite{FS}. 
Finally, in Fig.~\ref{peaketc}c we plot $\chi_4(t^*)$ \textit{vs.}
$t^*$. We find that slight deviations from power laws that are 
visible in Figs.~\ref{peaketc}a and \ref{peaketc}b are magnified in 
Fig.~\ref{peaketc}c.
Nevertheless we can still fit a power law $\chi_4(t^*) \propto 
(t^*)^{1/\gamma_2}$ with the exponent $\gamma_2=2.31 \pm 0.21$. 

\begin{figure}
	\includegraphics[scale=0.25]{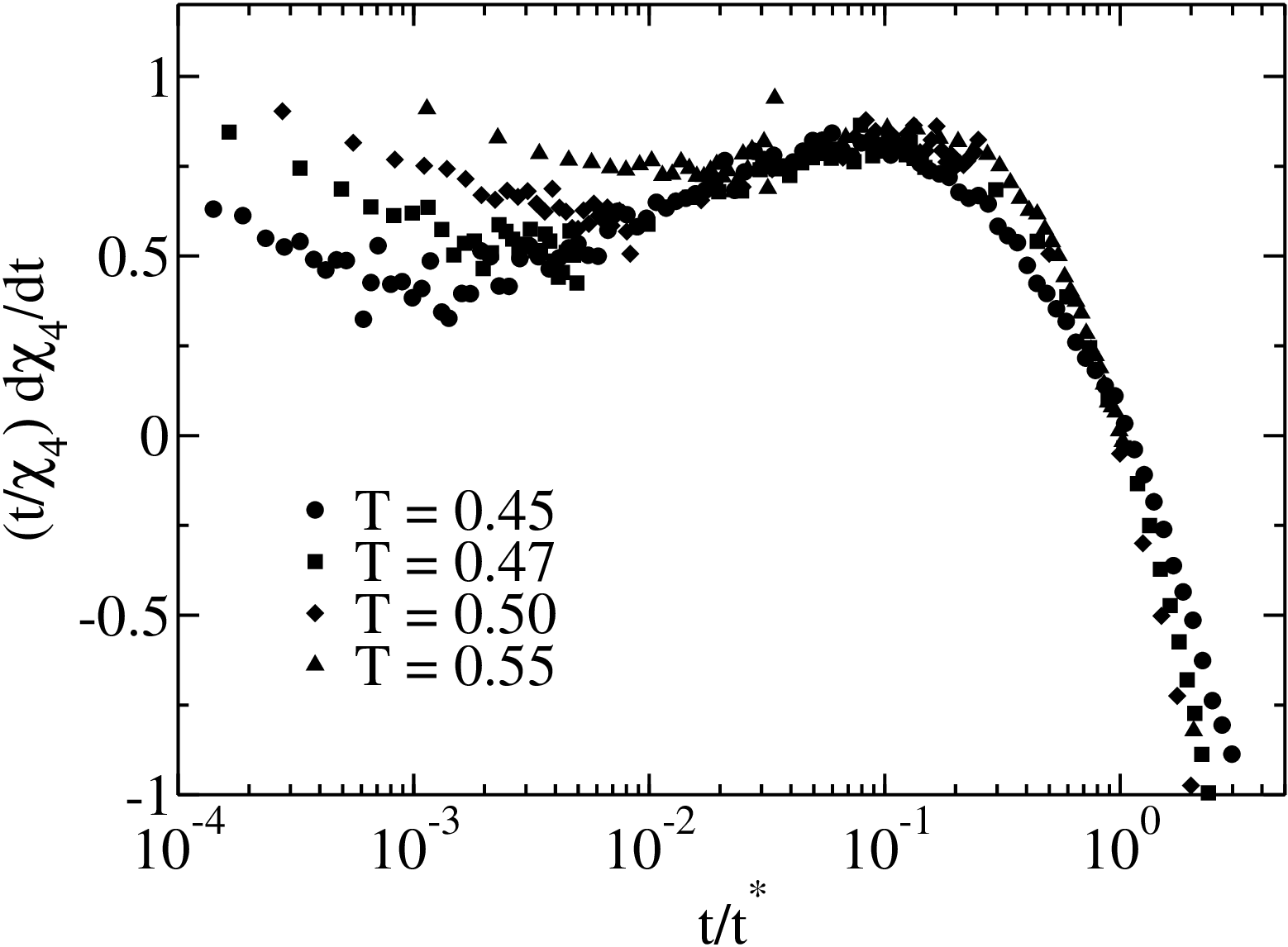}
\caption{\label{power} Derivative of $\ln \chi_4(t)$ w.r.t $\ln t$,
$d \ln \chi_4(t) / d\ln t = (t/\chi_4(t)) d\chi_4(t)/dt$,
plotted \textit{vs.} $t/t^*$ for $T=$ 0.45, 0.47, 0.50, and 0.55
(left to right).}
\end{figure}

In Fig.~\ref{power} we address the question of the power law dependence of
the four-point susceptibility on time. Instead of trying to fit power laws
to $\chi_4(t)$ over some specific time intervals, 
we have numerically calculated the 
derivative of $\ln \chi_4(t)$ with respect to $\ln t$, 
$d \ln \chi_4(t) / d\ln t = 
(t/\chi_4(t)) d\chi_4(t)/dt$. For any time interval over which $\chi_4(t)$
depends on $t$ in a power law fashion,  $d \ln \chi_4(t) / d\ln t$ should have
a constant value (a plateau). On the basis of the predictions of 
Ref.~\cite{BBEPL} we expect two different plateaus in our data.
A plateau in the early-$\beta$ regime and another one in the late-$\beta$ 
regime, \textit{i.e.} on approaching the maximum of $\chi_4(t)$ (additional
plateaus are predicted for shorter times, and we do not analyze them here). 
However, the simulations agree with mode coupling predictions
only over a restricted range of temperatures. It is not clear whether 
these two time scales are well separated over this temperature range, 
thus it is not clear if we could see two different plateaus. 

Indeed, Fig. \ref{power} does not show well developed plateaus. 
However, we notice that $\chi_4(t)$ grows with $t$ with an exponent of 
approximately $b = 0.8$ upon approaching it's maximum for all
temperatures. 
This exponent is comparable to the exponents 
obtained from Newtonian dynamics simulations
\cite{Toninelli4p}. It is quite a bit higher than
the exponent obtained from the mode coupling theory $b^{th} = 0.62$
\cite{bremark}.
  
\section{\label{estimate}Test of the estimate of the four-point susceptibility
proposed by Berthier \textit{et al.}}

The four-point susceptibility can be easily obtained from 
computer simulations, but it is not readily determined from experiments.
To address this problem Berthier \textit{et al.}~\cite{BerthierScience}
proposed an approximate estimate for the four-point susceptibility 
in terms of the derivative of the self-intermediate scattering function
with respect to temperature, $\chi_T(t)$,
\begin{equation}\label{chiT}
\chi_T(t) = \frac{\partial F_s(k;t)}{\partial T}.
\end{equation}

The starting point of their argument was a fluctuation-dissipation relation
(Eq. (2) of Ref. \cite{BerthierScience}).
If we naively adopted this relation to our system 
(\textit{i.e.} Brownian dynamics, NVT ensemble) we would get 
\begin{equation}\label{FDT}
k_B T^2 \chi_T(t) = 
\left< \delta F_s(\vec{k};t) \delta V(0) \right>.
\end{equation} 
Here $\delta V(t)$ denotes the instantaneous fluctuation of the total
potential energy,
\begin{equation}\label{V} 
V(t) = \sum_{i,j}\!'\sum_{\alpha\beta} 
V_{\alpha \beta}\left(\left|\vec{r}\,_i^{\alpha}(t) - 
\vec{r}\,_j^{\beta}(t) \right| \right),
\end{equation} 
\begin{equation}\label{deltaV} 
\delta V(t) = V(t) - \left<V(t)\right>.
\end{equation} 
The derivative with respect to the temperature in
Eq.~(\ref{estimate}), Eq.~(\ref{FDT}) and in the remainder of this section 
have to be calculated while keeping the short-time diffusion
coefficient $D_0$ constant since our short-time diffusion 
coefficient is proportional to the temperature. 

The main analytical result obtained by Berthier \textit{et al.} was an exact 
lower bound for $\chi_4(t)$ in terms of $\chi_T(t)$
(Eq. (5) of Ref.~\cite{BerthierScience}). Naively applying this 
result to our Brownian system we would get
\begin{equation}\label{bound}
\chi_4(t) \ge \frac{k_B}{c^{pot}_V} T^2 \chi_T^2(t).
\end{equation}
Here $c^{pot}_V$ is the potential contribution to the constant volume
specific heat per particle.

Berthier \textit{et al.} found that the difference
between the right and left sides of relation
(\ref{bound}) diminishes with decreasing temperature. On this basis, they 
proposed using the right-hand-side of Eq.~(\ref{bound}) as an approximate 
estimate for the four-point susceptibility. Furthermore, they showed that this
estimate can be easily calculated using either experimental or simulational
results for the self-intermediate scattering function. 

\begin{figure}
	\includegraphics[scale=0.25]{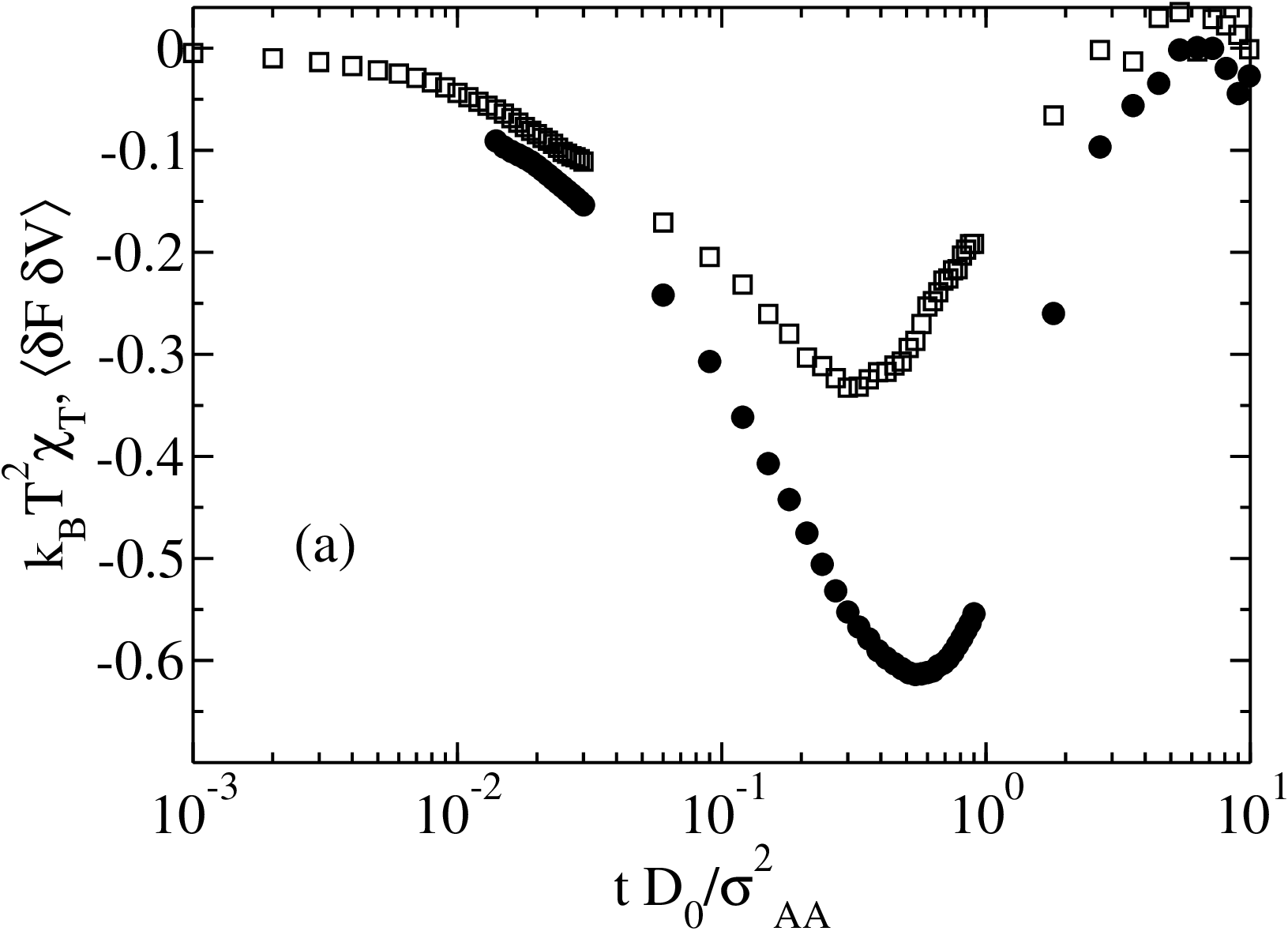}\\[0.25cm]
	\includegraphics[scale=0.25]{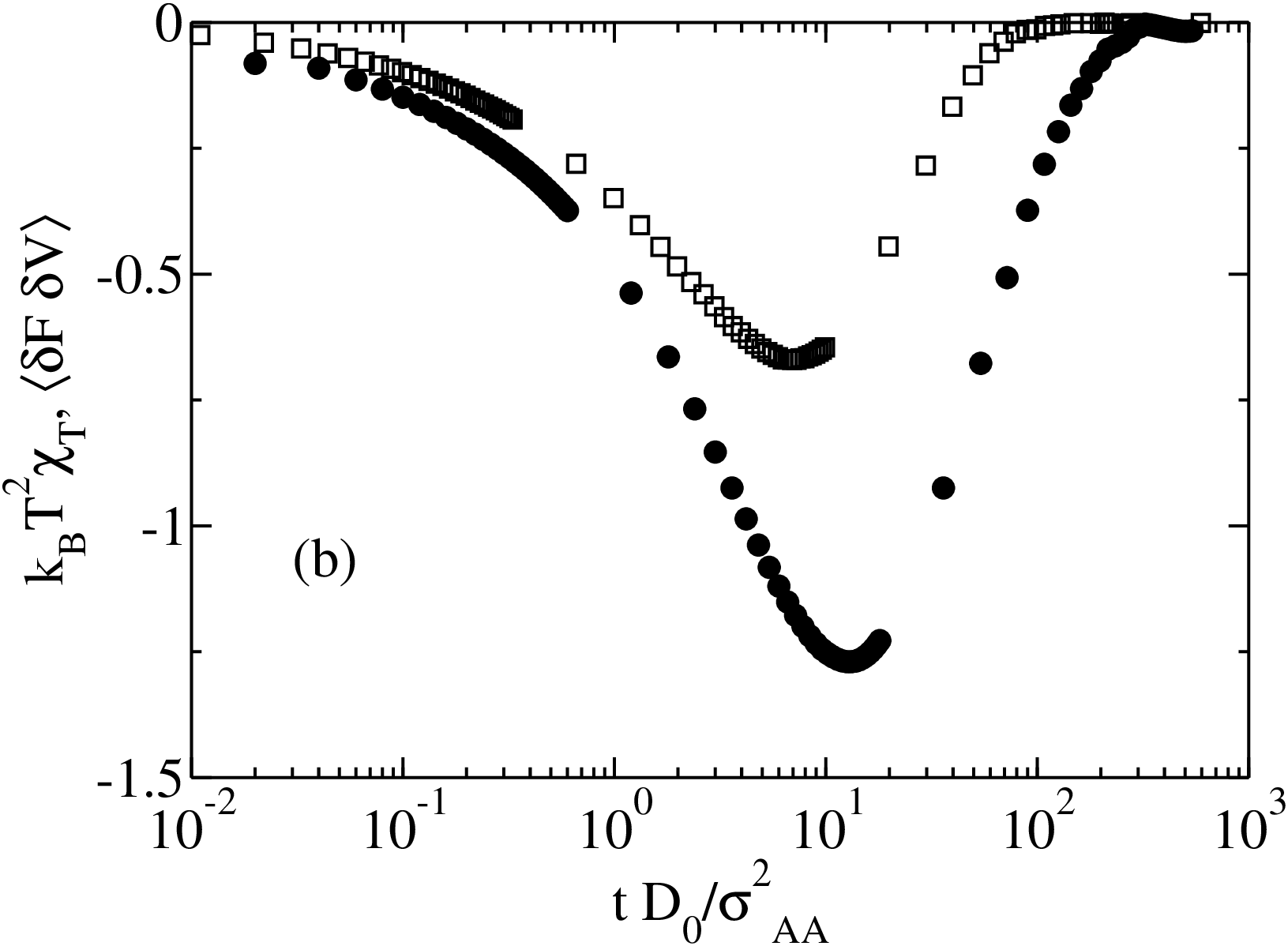}\\[0.25cm]
	\includegraphics[scale=0.25]{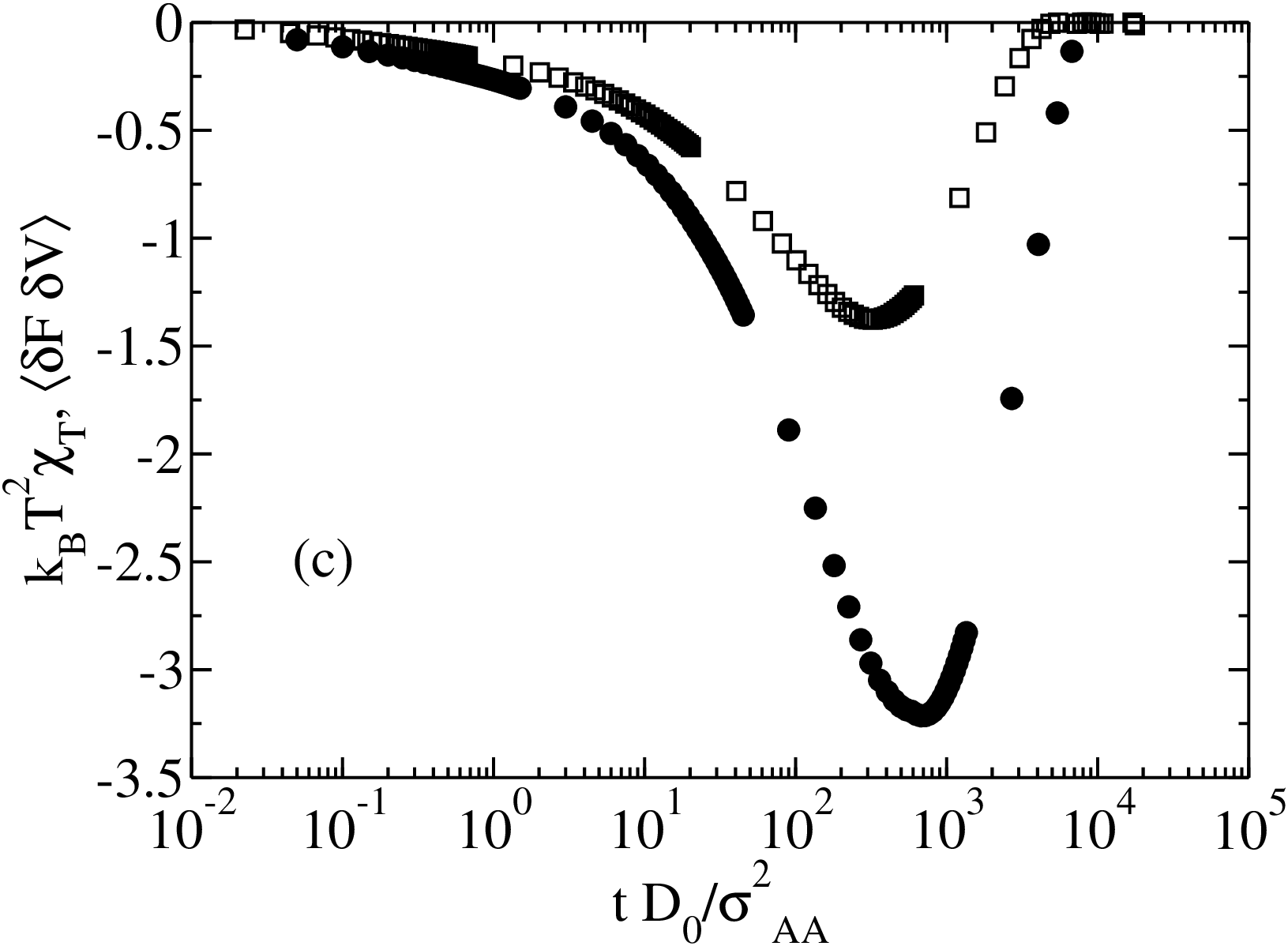}
\caption{\label{chiTcomp}
Comparison of $k_B T^2 \chi_T(t)= k_B T^2 \partial F(k;t)/\partial T$ 
(closed circles) with  $\left< \delta F_s(\vec{k};t) \delta V(0) \right>$
(open squares) at three representative temperatures: 
(a) T=1.00,
(b) T=0.55 and 
(c) T=0.45.}
\end{figure}

\begin{figure}
	\includegraphics[scale=0.25]{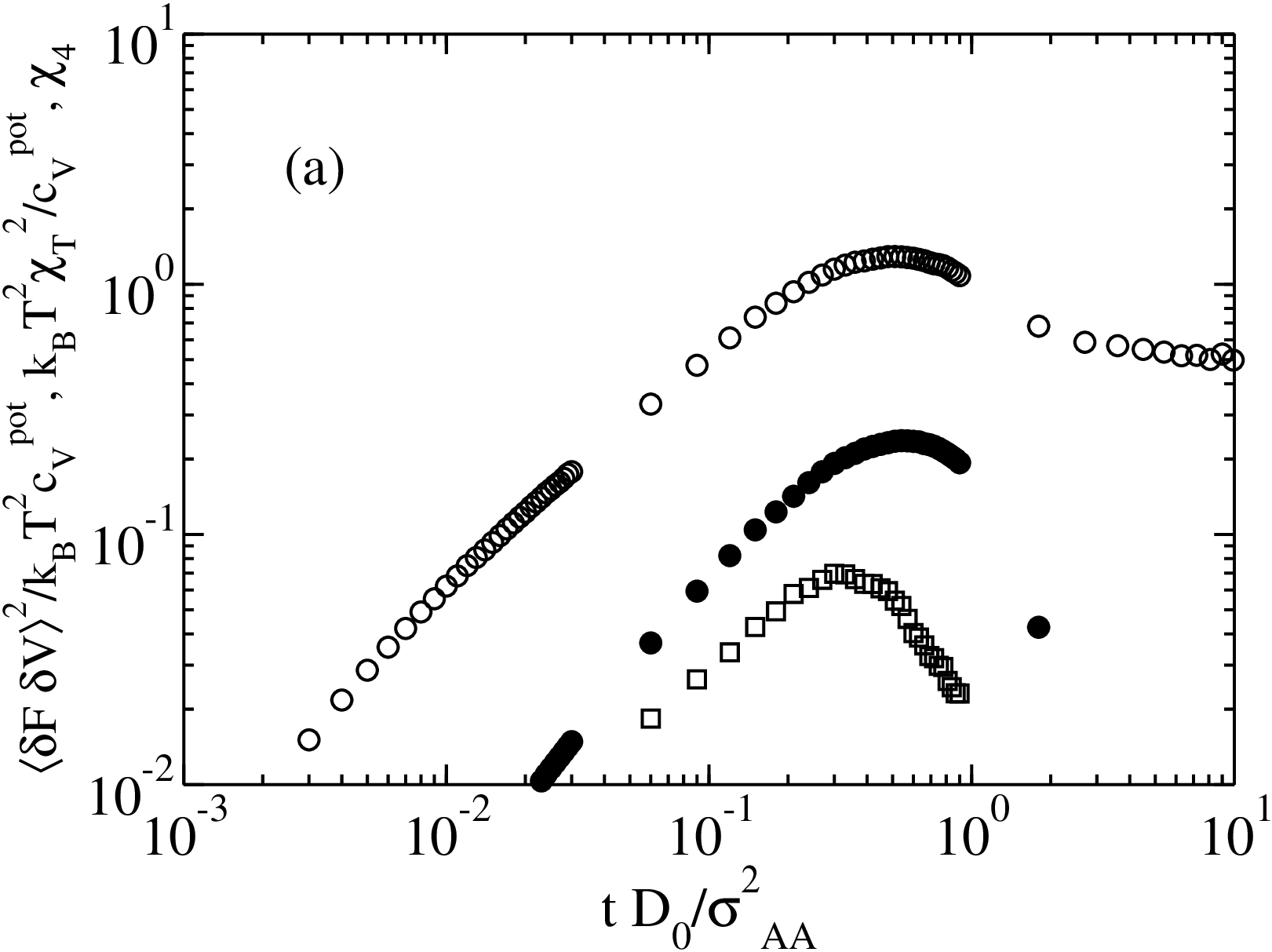}\\[0.25cm]
	\includegraphics[scale=0.25]{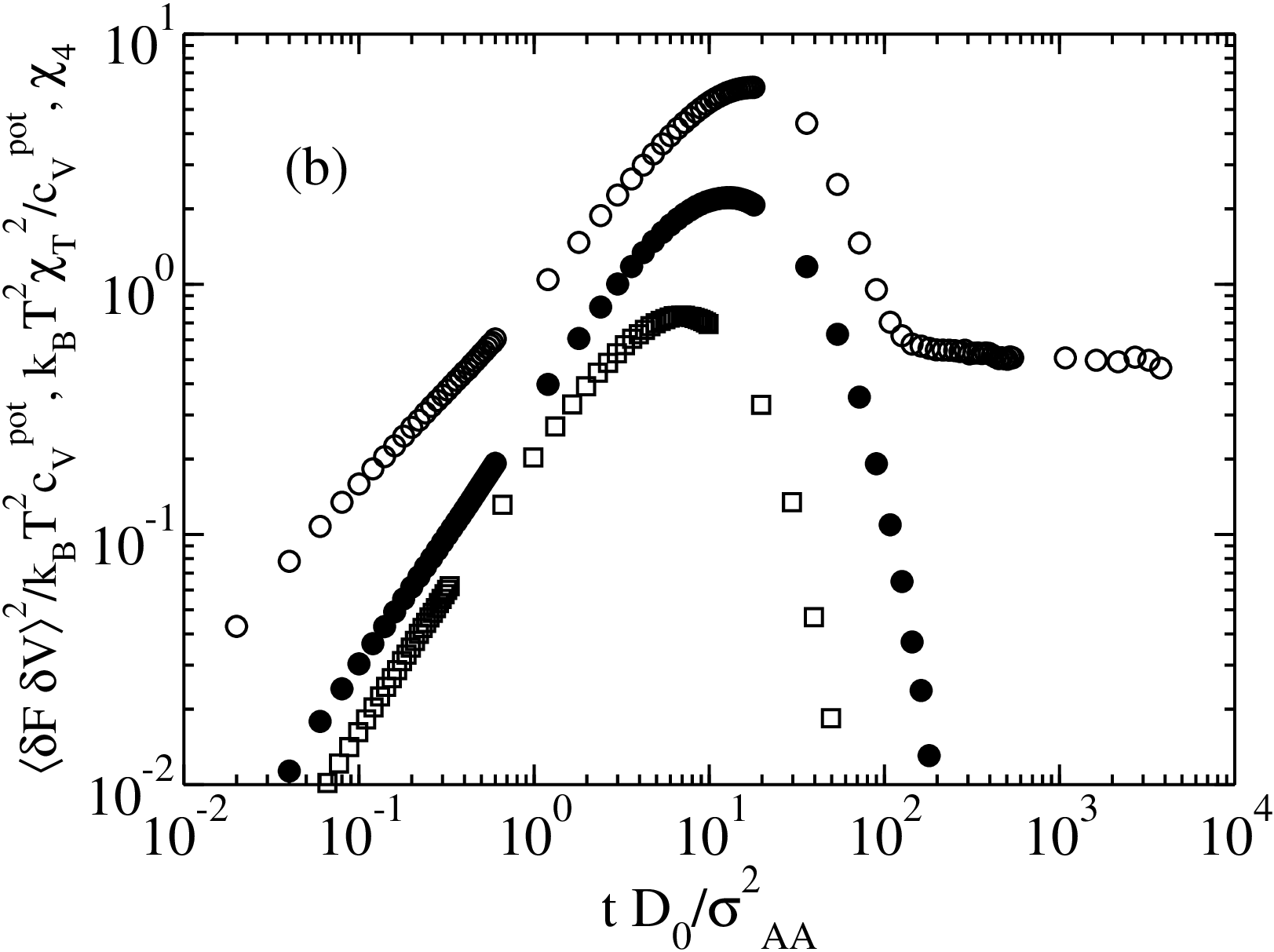}\\[0.25cm]
	\includegraphics[scale=0.25]{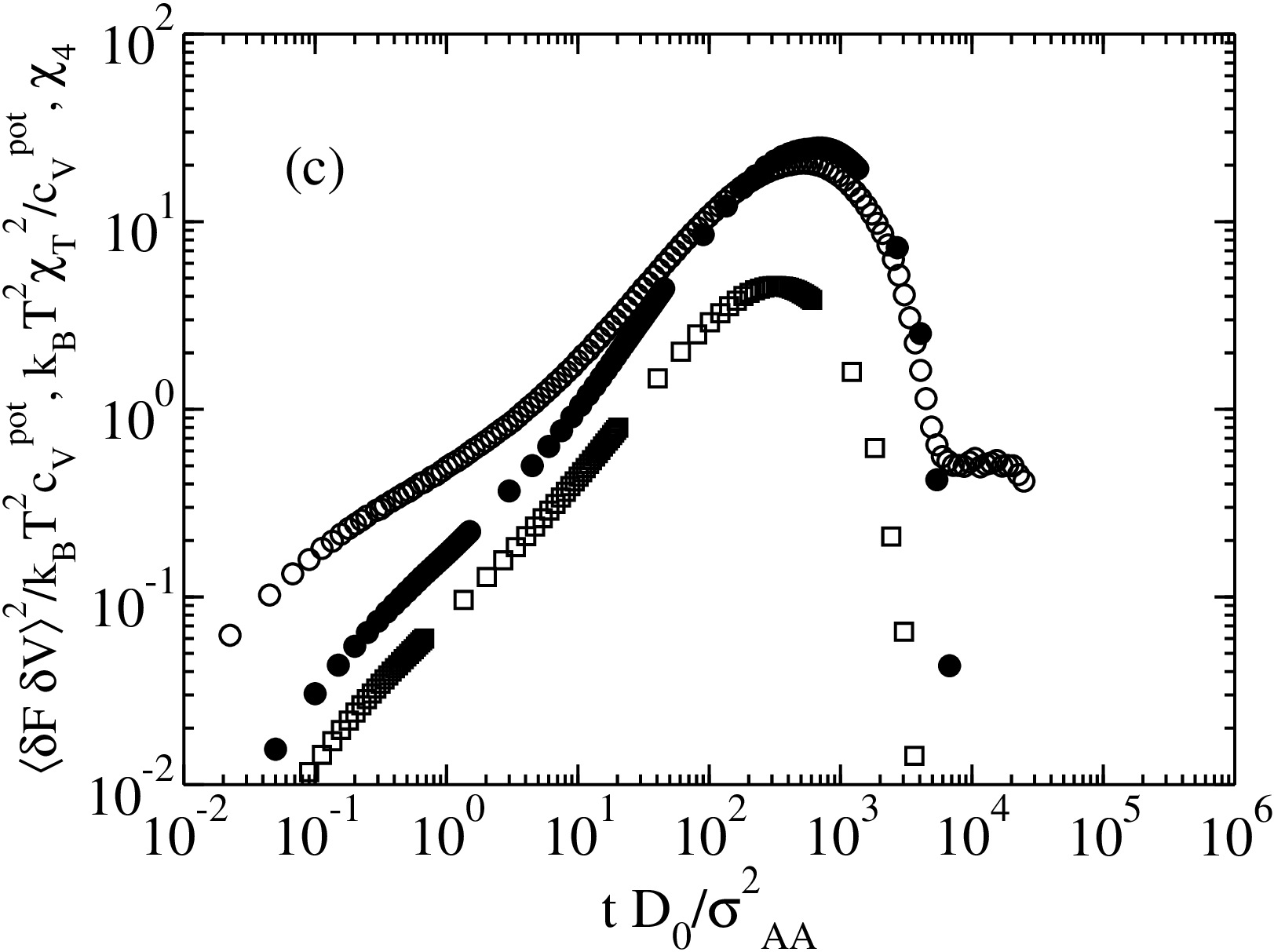}
\caption{\label{chi4comp}
Comparison of $\chi_4(t)$ (open circles) with 
$k_B T^2 \chi_T^2(t)/c_V^{pot}$ 
(closed circles) and  
$\left< \delta F_s(\vec{k};t) \delta V(0) \right>^2/(k_B T^2 c^{pot}_V)$
(open squares) at three representative temperatures: 
(a) T=1.00,
(b) T=0.55 and 
(c) T=0.45.}
\end{figure}

Both the fluctuation-dissipation relation 
and the bound derived by Berthier \textit{et al.}~are valid only
if the equations of motion do not involve the 
temperature. The only place where the temperature can enter is the 
initial condition which is given by the canonical distribution. 
This is true neither for the usual NVT computer simulations,
where the temperature enters into the equations of motion 
\textit{via} a thermostat, nor for our Brownian dynamics 
simulations, where the temperature enters \textit{via} the noise strength. 
Thus we have no arguments in favor of either the 
fluctuation-dissipation relation (\ref{FDT}) or the bound (\ref{bound}) for
our system.
In fact, we show explicitly in Fig. \ref{chiTcomp} that the relation
(\ref{FDT}) is violated in Brownian systems. On the other hand
we do not have strong numerical evidence for the violation of
the inequality (\ref{bound}). Figure \ref{chi4comp}c suggests that
this inequality is violated, but the extent of the violation is smaller
than the error bars. 

We should mention that there is a related, exact bound for the four-point 
susceptibility that follows from the Cauchy-Schwarz inequality 
(see footnote 22 of Ref. \cite{BerthierScience}):
\begin{equation}\label{boundCS}
\chi_4(t) \ge \frac{\left< \delta F_s(\vec{k};t) \delta V(0) \right>^2}{k_B T^2 c^{pot}_V}.
\end{equation}
However, as we show in Fig. \ref{chi4comp} this bound does not lead
to a useful estimate for the four-point susceptibility.

Berthier \textit{et al.} also gave a different, general 
argument leading to an estimate for the four-point susceptibility in terms of
the right-hand-side of inequality (\ref{bound}).
This argument can be easily adopted for our system.  If we assume that 
the main source of fluctuations of the instantaneous expression for the
scattering function is the potential energy, we get
\begin{equation}\label{fluct}
\delta F_s(\vec{k};t) \approx 
\frac{\partial F_s(k;t)}{\partial T}
\frac{\delta V(0)}{N c^{pot}_V}.
\end{equation}
Eq. (\ref{fluct}) leads immediately to 
\begin{equation}\label{est}
\chi_4(t) \approx \frac{k_B}{c^{pot}_V} T^2 \chi_T^2(t).
\end{equation}

In Fig. \ref{chi4comp}
we show the time dependence of both sides of the 
relation (\ref{est}) at three representative temperatures. 
While the approximation (\ref{est}) is inaccurate,
it becomes better, especially near the maximum value of $\chi_4(t)$,
when the temperature approaches the mode coupling temperature.

\begin{figure}
	\includegraphics[scale=0.25]{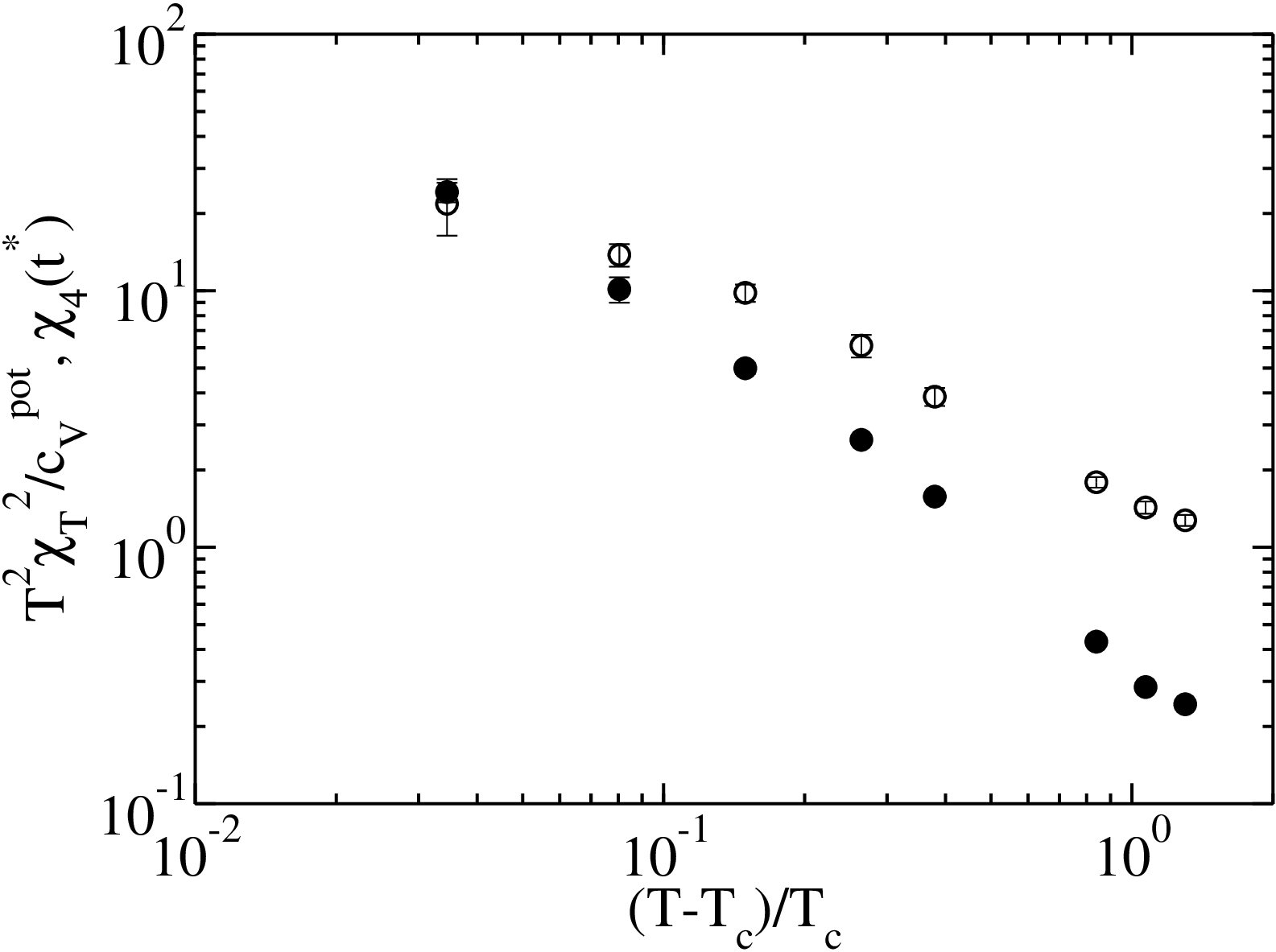}
\caption{\label{peakcomp} Temperature dependence of the maximum value of 
$k_B T^2 \chi_T^2(t)/c_V^{pot}$ 
(closed circles) compared to that of the maximum value of $\chi_4(t)$
(open circles).}
\end{figure}

Figure \ref{peakcomp} compares 
the temperature dependence of the maximum value of the
right-hand-side of the relation (\ref{est}) to the temperature
dependence of the maximum value of the  
four-point susceptibility. The difference between the two diminishes
with decreasing temperature. However, it is clear that 
the temperature dependence of these quantities is different. In particular,
one should be cautious when using
the temperature dependence of the estimate (\ref{est}) to obtain the
temperature dependence of the dynamic correlation length, at least for
temperatures above the mode coupling temperature.

\begin{figure}
	\includegraphics[scale=0.25]{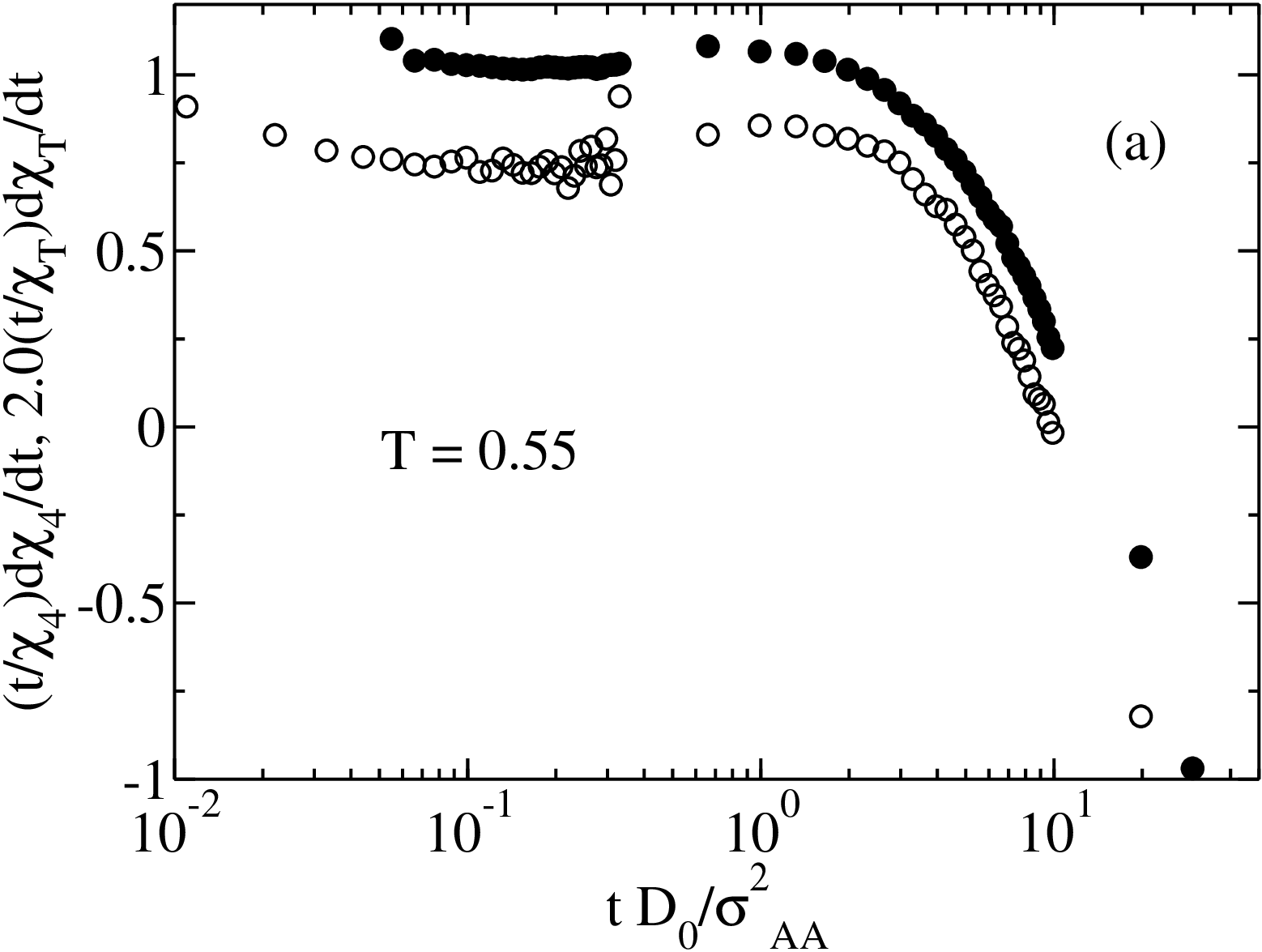}\\[0.25cm]
	\includegraphics[scale=0.25]{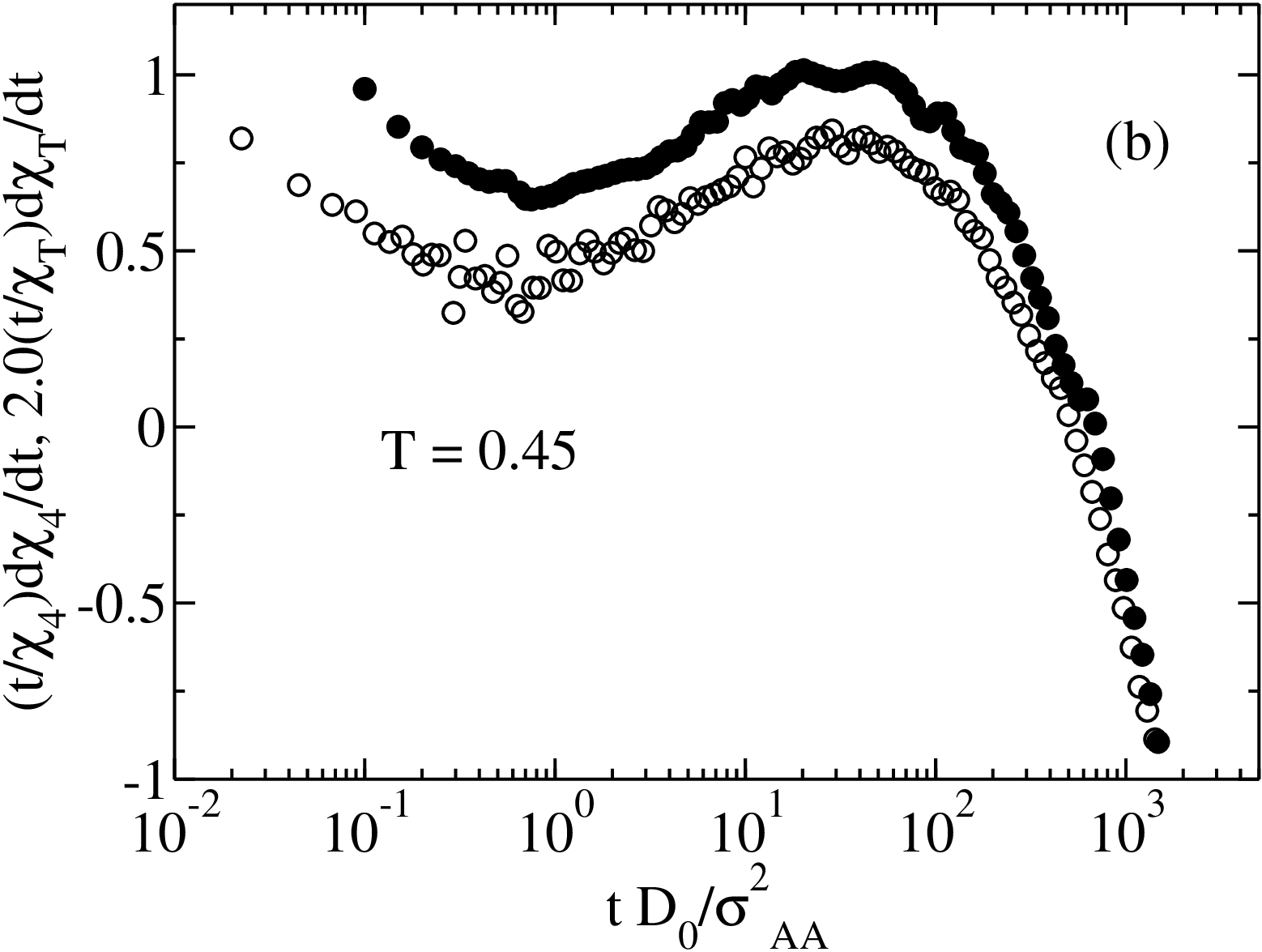}
\caption{\label{chi4dercomp}
Comparison of the derivative 
$d\ln \chi_4(t)/d \ln t = (t/\chi_4(t))d\chi_4(t)/dt$ 
(open circles) and the derivative 
$d\ln k_B T^2 \chi_T^2(t)/c^{pot}_V/d\ln t = 2(t/\chi_T(t))d\chi_T(t)/dt$ 
(closed circles) at two temperatures: (a) T=0.55, and (b) T=0.45.}
\end{figure}

Finally, Fig. \ref{chi4dercomp} compares the derivative of the logarithm 
of the four-point susceptibility, $\ln \chi_4(t)$,  
with respect to the logarithm of time, $\ln t$
and the derivative of the logarithm of the estimate (\ref{est}), 
$\ln k_B T^2 \chi_T^2(t)/c^{pot}_V = 2\ln \chi_T(t) + const.$, 
with respect to $\ln t$. We find that the estimate 
(\ref{est}) approaches its maximum value with a power law exponent that 
is approximately equal to 1 and thus somewhat higher than the power
law exponent describing $\chi_4(t)$'s approach to its maximum value. 
Moreover, comparing Figs. \ref{chi4dercomp}a and \ref{chi4dercomp}b 
we find that the
difference between exponents obtained from the estimate (\ref{est}) 
and from the four-point susceptibility
does not seem to diminish with decreasing temperature. 

\section{\label{conclusion}Conclusions}

We performed a quantitative analysis of the four-point susceptibility
$\chi_4(t)$ of the Kob-Andersen Lennard-Jones binary mixture. We
compared the results of Brownian dynamics computer simulations 
to predictions obtained from a recent re-formulation of the  
mode coupling theory. We did not compare computer simulation results 
to other approaches to glassy dynamics (\textit{e.g.}, to the predictions 
obtained using facilitated kinetic Ising models \cite{fkIm1,fkIm2}) 
because these 
other approaches are more concerned with a temperature range 
lower than the one accessible in our simulations.

We found that some of the mode coupling predictions agree with our simulation
results. Most notably, the height of the maximum of the four-point 
susceptibility grows upon approaching the mode coupling temperature from above
in the way predicted by the theory. Moreover, the time at which $\chi_4(t)$
reaches its maximum value has the same temperature dependence as the
$\alpha$ relaxation time. For each of these two quantities, 
which are derived from 
a \emph{four}-point function, the power law  dependence on
$(T-T_c)/T_c$ is obeyed over the same temperature range as for quantities 
derived from \emph{two}-point functions, \textit{e.g.}
the $\alpha$ relaxation time and the self-diffusion coefficient.

On the other hand, we found some disagreement with the theory with
regard to the power law dependence of $\chi_4(t)$ on time. 
We found that upon approaching its maximum value $\chi_4(t)$ grows with
time with an effective exponent of about 0.8.
This exponent is quite a bit larger than the exponent predicted by the theory.
It should be recalled that the exponent predicted by the theory is
in turn larger than the one obtained by fitting a formula inspired by
the mode coupling theory to the (two-point) 
self-intermediate scattering function. Hence, our understanding of the 
connection between the time dependence of the four-point susceptibility 
and the time dependence of two-point functions seems incomplete.

Finally, we tested the approximate estimate of the four-point susceptibility
in terms of the temperature derivative of the self-intermediate scattering
function. We found that
the estimate becomes accurate around the peak of $\chi_4(t)$ upon approaching
the mode coupling temperature. However,
the temperature dependence of the maximum value of $\chi_4(t)$ is 
weaker than that of the approximate estimate. Moreover, the time
dependence of $\chi_4(t)$ differs from that of the estimate even
near the mode coupling temperature. 

We should emphasize that all our computer simulation results were obtained
using Brownian dynamics (and, therefore, NVT ensemble). It is possible
that results obtained from Newtonian dynamics computer simulations 
would agree better (or worse) with theoretical predictions.

\section*{Acknowledgments}
We gratefully acknowledge the support of NSF Grant No.~CHE 0517709.


\begin{thebibliography}{99}
\bibitem{bremark} We obtained $b^{th}=0.62$ by fitting the asymptotic
functional form predicted by the mode coupling theory to the solution
of the full mode coupling equations \cite{FS}. This value is very close
to $b^{th} = 0.63$ quoted in \cite{Toninelli4p}.
\end{thebibliography}
\end{document}